# Powerful flaring configurations in solar active regions: magnetic topology via special features of $\delta$-structure phenomenon

## I. Basic principles of topological model

R.N.Ikhsanov and Yu.V.Marushin

Central Astronomical Observatory of the Russian Academy of Science
65-1 Pulkovo, St.Petersburg 196140, Russia, e-mail: `solar1@gao.spb.ru` *

**Abstract.** The close connection of magnetic field structure on the one hand and observable features of complex active regions with high flare efficiency on the other one is investigated within the framework of the topological model of interacting magnetic complexes (IMC) and dynamic classification of flaring magnetic configurations (FMC) suggested by Ikhsanov (1982). The primary objective of the present first part of this work is to expose general statements of the specified model and classification, the secondary one assumes the critical analysis and logical generalization of several important regularities which are frequently observable in evolution stages of a typical complex active region (AR) with powerful flares including proton ones. Those cases are basically examined where the development of photospheric situation was marked by the formation of $\delta$-configuration phenomenon described by Künzel (1960) as one of the most impressive precursors of high flare activity. Without trying to use any additional assumptions and alternative hypotheses, the qualitative analysis of physical reasons and formative conditions of various $\delta$-configuration types in such ARs is carried out. An attempt to construct the non-contradictory physical interpretation of morphological, flare-productive and magnetic features of such configurations within the framework of Ikhsanov's IMC model is also done. As a result of this generalization it was possible to show that the complex analysis of $\delta$-structure phenomenon from the IMC-based point of view allows to explain well many early established regularities as kindred phenomena having the uniform physical reason. It is obviously possible to approve in authors's opinion that the nontrivial evolution of the FMC of a typical active region accompanied by the formation of a number of new magnetic complexes and their further interaction with old field structures arisen earlier could be considered as such a reason. The authors intend to demonstrate that the observable consequences of this process seem to be vitally determined not only by capacities of magnetic fluxes themselves but also by a priority of their emergence from under the photosphere and by some special features of their mutual arrangement. The detailed analysis of the specified conditionality along with the discussion of the proposed typical ways to interpret main regularities through the IMC model are essential topic of the present work. The practical application of the model requires the consideration of extensive observational data concerning several complex ARs with powerful flares. It will be consistently carried out in subsequent second part of this study.

## 1. Introduction

The resolving of a problem of solar flares seems to be impossible without researches of the structure of magnetic fields along with observable regularities of their evolution during the life period of an active region. Really, the huge file of high precision observations of the chromosphere and corona saved up to the present time does not find out any indications on existence in an atmosphere of the Sun of any other energy sources allowing to explain a phenomenon of solar flare except for magnetic field. Therefore the most proved and standard assumption is considered now that the power of flares is mainly provided by electromagnetic process and the whole complex of secondary phenomena forming the observable picture of a flare is closely connected to a wide number of physical events occurring in AR magnetic field.

It is well known that the brightest and historically first manifestation of such fields from found out to the present time is sunspot formation activity having expressed *group* character. It is reliably established by magnetographic observations (see, for example: Martin & Ramsy 1972) that solar flares are frequently accompanied by essential reorganization of magnetic field structure within an active region. By one of possible reasons of such a reorganization the existing theory considers (see, for example: Křivsky 1968; Tomozov 1976; Priest 1976; Heyvaerts et al. 1977; Heyvaerts & Kuperus 1978) the emergence of a new magnetic flux from below the photosphere. Under certain conditions that one can be able to strongly alterate the pre-existing field topology and so result in quite rapide modification of AR magnetic configuration.

Being considered in outline the development of magnetic field structures resulting in flare phenomenon has its own history which can be logically separated into three basic stages, apparently. The first stage corresponds to the period of initial formation of AR magnetic field all along with its specific fea-

---

*Correspondence to*: R.N.Ikhsanov (`solar1@gao.spb.ru`)

* Please send a copy to Yu.V.Marushin (`uvmar@mail.ru`)

tures. The second one is the emergence of the field on a surface of the Sun including the development of a suitable active region and the formation of appropriate magnetic configuration. The nontrivial evolution of such a configuration next results in the third stage which conforms with the initiation and development of flare process of a certain capacity. The first stage is closed in essence with the problem of solar magnetic field generation. The third one concerns directly the mechanism of a flare and, hence, requires the strict knowledge of fine structure of a field within the limits of those spatial areas in the solar atmosphere which the flare is originated from. However, the preparation of AR magnetic field for flare realization always occurs during the second stage. Moreover, only that one is accessible for direct observations at present. Therefore it is possible to approve that the analysis of magnetic topology evolution carried out along with the consideration of probable reasons and mechanisms causing its fast complication and further relaxation is the necessary point to understand two other stages, specifically the third one.

For this reason the following observational tasks should be considered as most important stages on the way to understand the mechanism of solar flare: (i) the detailed consideration of magnetic field structure in sunspot groups and (ii) the search for any regularities in the formation of the most typical magnetic configurations within the flaring region. Certainly, main interest represent such magnetic configurations which the development is accompanied by *powerful* flares and, essentially, those ones where a level of flare efficiency is empirically *known*.

It is necessary to note that the understanding of the fact that flare phenomena itself is closely connected to the physics of solar magnetic fields finds its reflection in modern practice of flare forecasting when various AR parameters directly or indirectly reflecting peculiarities of magnetic field structure are widely used as flare precursors. However, unfortunately, not only the problem to determine a set of conditions which appear to be necessary and sufficient for the occurrence of flares themselves but *even* the question on strict formative conditions of such *precursors* is still too far from its final resolution and now represents significant interest.

It is no mere chance that the morphological analysis of those distinctive features of sunspot group structure which find out some properties of high flare activity precursors is very important topic that heliophysics traditionally carry to a number of key directions in flare forecasting. As the formation of $\delta$-configuration phenomenon in complex sunspot groups is one of the most impressive precursors of a similar kind, we can under right to consider $\delta$-structure researches to be a certainly urgent scientific problem.

Really, it was marked for the first time by Künzel more than forty years ago (Künzel 1960) that most powerful flares have the curious but obvious tendency to occur in complex sunspot groups having characteristic morphological peculiarity, which he has named as $\delta$-*configuration* . As known, the specified feature consist that two or more regions of *opposite* polarity were located in all such groups within the limits of one *common* penumbra. The subsequent works have shown that the fact of $\delta$-structure formation in this or that group not yet guarantees the compulsory increase of its flare potential. In other words, not each $\delta$-configuration is flare-effective (Künzel et al. 1961). Unfortunately, no physical explanations of observable properties of $\delta$-structures as powerful flare precursors was suggested by Künzel. As a result, the phenomenon remained an object of clean empirical and statistical analysis for a long time.

However, numerous confirmations of an electromagnetic nature of physical processes in the active regions, for the first time reliably received in 60's, have resulted in understanding of the fact that even the clean morphological phenomena in flare-productive sunspot groups can and *must* be always interpreted in the aggregate with the analysis of magnetic field features. Such a consideration is now carrying out within the framework of various author's approaches. So, for example, since a moment of Künzel's work (1960) appearance the study of $\delta$-configurations have turned into one of the rather important points in heliophysics (see, for example: Smith & Howard 1968) and continue to remain an object analysing by many authors (see, for example: Zirin & Liggett 1987). In this context the problem to find out some optimum conditions for the formation of $\delta$-configurations of various types as well as the construction of adequate physical interpretation of their observable features is also rather actual.

The correct clarification of this subject assumes an analysis of extensive observational data for a number of complex active regions of various types allowing to look in details after the development picture of each one and, specifically, to detect the emergence of their magnetic fields on a surface of the Sun. The results of author's attempt of such analysis with reference to $\delta$-structure phenomenon are based on the other qualitative approach, namely on the topological model of interacting magnetic complexes (hereafter the IMC model) in the aggregate with the so-called dynamic classification of flaring magnetic configurations (hereafter the FMC classification). The first part of results to be presented only concerns the most common questions of $\delta$-structure formation in a typical complex AR with high flare efficiency. This includes the critical analysis of the IMC model and the qualitative demonstration of how the non-contradictory interpretation of a number of observable regularities in such ARs can be constructed on this basis. The detailed exposition of all these topics is the subject of the present work.

## 2. IMC topological model and FMC dynamic classification

First of all, it is necessary to remember that the heliophysics is traditionally very attentive to the problem of a new magnetic flux emergence as well as to the study of appropriate structures of magnetic fields, so that a number of excellent reviews is published on this problem to the present time (see, for example: Kurokawa 1991). Unfortunately, at the same time main rules of the IMC topological model and FMC dynamic classification (Ikhsanov 1982) were not published for some reasons in western periodical press before now. Certainly, we shall not discuss these reasons here, instead of this we shall only note that the specified circumstance requires us to be briefly stopped hereafter on the key moments in a history of researches of this question.

## 2.1. Magnetic flux emergence and solar flares

The initial point of researches became Parker's work (Parker 1955) where it was theoretically shown that the horizontal magnetic flux tube must be unstable, and so will tend to rise toward the surface of the Sun being influenced by the mechanism called "magnetic buoyancy". The reason of the last effect is the presence of magnetic pressure inside a flux tube containing a field, so that the buoyancy phenomenon itself occurs as a result of gas pressure inequality inside and outside the volume element which must lead in locally isothermal conditions to "arhimedian" inequality of densities.

Further in 1964–1966 Vitinsky and Ikhsanov (1964) have shown on extensive observational data that the usual "scatter" motion of a pair of main sunspots within several bipolar groups seems to be explained best by the presence of a new magnetic flux tube. Following the authors, the last one takes a semitorroidal shape and rises from under the photosphere with the velocity of about $105 \pm 30$ m/s. It was also shown that such a tube does not leave back under the photosphere but dissipates within the higher layers of solar atmosphere as a rule (Vitinsky & Ikhsanov 1966). Moreover, it was found that the capacity of a new bipolar complex (i.e. approximately the areas of the leading and following sunspots) as well as the size of a tube and the duration of its rise are closely related among themselves. Thus, Parker's theoretical opportunity of a new magnetic flux emergence from under the photosphere which is next observable as sunspots, plages etc and further brings to diffusion of interacting field structures into higher layers of solar atmosphere has received an independent experimental confirmation in addition to theoretical substantiation. Hereinafter, the rise of magnetic field structures repeatedly proved to be true by various observational methods and now starts as an established fact.

A little before the series of works carried out in Crimean astrophysical observatory in the beginning of 60's have revealed for the first time (Severny 1958, 1960; Moreton & Severny 1968) that the flares are usually grouped near to neutral line $H_\parallel = 0$ of AR magnetic field. What's usual for sunspot groups producing proton flares that this line is frequently located along the direction $E - W$ (Zvereva & Severny 1970) or takes a distinctive "zigzag" shape. The fact that the gradient of longitudinal magnetic field $\Delta B_\parallel / \Delta r$ usually increases before flares was also reliably established and it was also marked that this value exceeds $10^{-1}$ G/km in the cases of powerful flares (Gopasyuk et al. 1963; Avignon et al. 1964, 1966; Godovnikov et al. 1964; Zvereva & Severny 1970). At the same time the value of magnetic energy $|W|$ reaches its maximum about $10^{32}$ erg allowing to explain a phenomenon of such flares from the quantitative point of view (Zvereva & Severny 1970).

At the same time Ogir and Shaposhnikova (1965) have found that sunspot formation is frequently observed before flares, whereas other authors marked an occurrence of sunspot satellites (see, for example: Kassinsky 1976). The variations of longitudinal $B_\parallel$ and transverse $B_\perp$ field component strengths derived from magnetographic measurements for some time before flares have also confirmed an emergence of a new magnetic flux allocated *near* or *inside* the old one and have shown that this flux had more often an *opposite* polarity (Ikhsanov 1974; Rust 1976).

## 2.2. Interaction of emerging fluxes: magnectic complexes and configurations

The following important stage became Ikhsanov's work (1974) where the author's attention was first drawn to the fact that one powerful flare in sunspot group has arisen in rather unusual situation. The marked specificity consist that two magnetic flux ropes have appeared to be "crosswise overlapped" in such a manner that their emergence has taken place directly under one another, whereas magnetic axes of the upper and lower systems were *revolved* concerning each other by a significant angle up to $90°$. In this case the tops of two magnetic arches come nearer to each other while rising from under the photosphere, so that the transverse field $B_\perp$ directions demonstrate either their cross-shaped intersection or at least the situation where they turn in due course and with changing the height. Here it is important to note the fact that the phenomenon of such an intersection (so-called "bifurcation") was first observed in itself on magnetographic data by Severny (1964), next by Zvereva and Severny (1970). However, its topological interpretation was first given and its confirmation was first received much later (Ikhsanov & Schegoleva 1980a, 1982) when a number of complex sunspot groups was closely considered.

It is also necessary to remind about the fact that already in 60–70's the phenomenon of a new magnetic flux emergence from under the photosphere was frequently observed as sudden formation of new sunspots (so-called "satellites") within the limits of existing sunspot groups. As a result, the close connection of this effect with solar flare occurrences was repeatedly marked by many authors (see, for example: Kassinsky 1976); at the same time first theoretical researches of different kinds of magnetic configurations were also carried out (Heyvaerts et al. 1976; Kaplan et al. 1977). However, the essentially important statement that the flare can be related to a certain *topology* of emerging fluxes, namely with a "crosswise overlapping" of two magnetic ropes arisen successively or simultaneously has been formulated for the first time in work (Ikhsanov 1974), as appear. Moreover, it has been also marked there that the *distinctive* feature of a typical active region with *powerful* flares is an interaction of *at least two* different emerging fluxes, where each one represents a separate system of magnetic ropes and belongs to an appropriate magnetic bipolar structure. From this point of view each flux could be considered as independent structural unit, so-called "magnetic complex", whereas the presence of two and more interacting magnetic complexes (along with specific features of their structure and their mutual arrangement) allows to speak about the formation of so-called "magnetic configuration".

## 2.3. Dynamic classification of magnetic configurations

Certainly, as the following stage of researches an attempt to analyse some real magnetic configurations of flaring regions was undertaken (Ikhsanov 1982). In order to obtain the detailed data on magnetic field structures for these regions, the "Catalogue of Sunspot Magnetic Fields during JGY period (1957–1958)" (Stepanov et al. 1963) along with the appendix to Solnechnye Dannye bulletin published in "Magnetic Fields of Sunspots" (1965–1974) were widely used. Simultaneously, the "Catalog of Solar Particle Events 1955–1969" (Svestka & Simon 1975), the "Quarterly Bulletin of Solar Activity" (1968–1969) and the works (Dodson & Hedeman 1960, 1968–1975) have also become the general sources where all available data on registered solar flares for the specified period were taken from.

As a result, a number of different magnetic configurations with rather simple topologies was selected from the whole variety of observable field structures as the most typical and frequently registered cases. On this basis the term *"flaring magnetic configuration"* (FMC) was formally introduced into the use and the topological formalism of the FMC classification is offered (Ikhsanov 1982). On the one hand, this has allowed to distinguish such configurations between them by analyzing the type of spatial interaction of two and more magnetic complexes emerging from under the photosphere. On the other one, by imposing certain topological restrictions on the structure of such interactions the FMC classification has given an opportunity to formalize the description problem of the variety of interaction schemes usually seen in magnetic complexes of solar flaring regions by restricting that one to within a topological class.

The simple idealized model forming the basis of the FMC classification widely uses schematic representation of real magnetic field structure of cleanly bipolar sunspot group as a unique magnetic arch or a tube. Really, according to early researches of sunspot thin structure (Ikhsanov 1972, 1973) each magnetic flux tube actually consists of a number of elementary magnetic ropes having the diameter of about $10^3$ km and incorporated into this or that hierarchical system of scales. Their stability is mainly ensured by the presence of a torsion and tension forces (Ikhsanov 1974). A set of several ropes having the common origin and resulting in a simple bipolar structure of a field was named as *"a complex of magnetic flux ropes"* or briefly *"a magnetic complex"*. At the points of intersection with the photosphere such a complex forms an appropriate bipolar pair of sunspots. Besides, if just a *single* sunspot is observed or there are several sunspots of *the same* polarity, this idealized model always considers such a situation as though there was just as well *the second* crossing of the photosphere by an appropriate magnetic rope but *without* any sunspot formation (Vitinsky & Ikhsanov 1966).

According to this scheme it is possible to consider sunspot group development to be determined by the emergence of one or several *magnetic complexes* from under the photospheric layers. Thus, the spatial position of a new complex concerning another seems to be the major factor determining the formation of flaring configurations within the magnetic field structure. Certainly, real magnetic complexes essentially can differ in their capacity i.e. in observable sizes of their magnetic arches. But if we intentionally simplify the situation and try to abstract from the total amount of those magnetic complexes which have arisen before (that is to accept them all for one) then the practice demonstrates that only *five* different FMC classes are more frequently observed during the rise of a new complex, not including a bipolar case. The corresponding magnetic topologies for all these classes along with their symbolic numbers and empirical definitions are summarized in Table 1 which the data is compiled in accordance with early researches (Ikhsanov 1982, Fig.1 and §1 therein).

As shown herein, the FMC classification assumes by default that the sizes of old and new complexes within the configurations of all classes are approximately identical, except for the class *IV*. Actually, the unequal sizes of old and new complexes can be observed, certainly, not only at the class *IV* but also at all other classes of magnetic configurations. Following the FMC classification this circumstance should be marked in such cases by addition of a letter *"m"* (this should equally mean *"mixed"* or *"modified"*) to the symbolic number of a suitable class. We shall specially note that there are sometimes more difficult cases where we have either (i) more than two magnetic complexes which interact in some non-trivial way but with a constant topology or (ii) the current interaction topology of two existing complexes varies in due course as a result of additional emergence of several new complexes along with the dissipation of old field structures within the top layers of solar atmosphere, or else (iii) both specified situations take place during all the period of sunspot group evolution or at its separate moments at any rate. In such difficult cases we has occasion either to attribute such configurations as topological objects of *several* classes *simultaneously* or to describe group evolution by making enumeration of all discovered classes in close conformity with the order of their formation as it was observed. Just due to the opportunity allowing not only to get the static characteristics of an "instant" field configuration by providing its qualitative description but also to describe and sometimes to predict its *changes* in process of an active region, the FMC classification has named *"dynamic"*.

### 2.4. Basic features of FMC dynamic classification

Without having an opportunity to discuss all numerous conclusions of Ikhsanov's work (1982) which could be considered here we shall only highlight the single statement which seems to be the most important point among others. It has turned out that the topological membership of a certain magnetic field structure in flare-productive AR to the FMC configuration of a certain class is connected by the *closest* way to observable features of its flare activity.

Particularly, it was shown on the basis of extensive observable data that with other things being equal the registered flare activity depends first of all on (i) relative arrangement of existing complexes and (ii) their magnetic polarity distribution, what means it is closely determined by a class of magnetic configuration as a first approximation. In addition to these reasons some other empirical factors can essentially affect flare productivity of a certain FMC configuration within the limits of its *own* class, namely (iii) geometrical distances between the sources of opposite polarities and also (iv) an absolute and relative capacity of interacting complexes. By using the term "rela-tive capacity" with reference to a pair of complexes we assume in the first place the real parity of cross-section diameters of appropriate magnetic tubes or just as well that one of the overall photospheric areas calculated for that pair of main sunspots which the specified tubes are "leaning" on.

**Table 1.** The formalism of FMC dynamic classification maintains six different topological classes. Real magnetic configurations usually seen in flaring ARs are well conforming with offered formal toplogies (see also Ikhsanov 1982, Table 1 herein).

| Dynamic class | Magnetic topology | Empirical class definition |
|---|---|---|
| I | 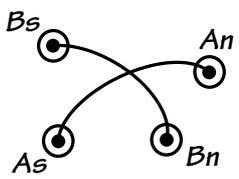 | A new magnetic complex arises *directly under* another old one with some displacement *on height*. Besides, the magnetic axes of a new lower complex (i.e. arisen later) and an old upper one (i.e. arisen before) are rotated regarding each other so that there is an angle up to $90°$ between them. This produces a distinctive phenomenon of *3D crosswise overlapping* of two magnetic flux ropes (also known as *bifurcation*) along with its observable consequences |
| II | 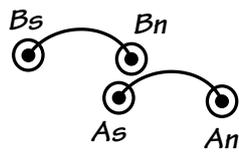 | The symmetry centers of two magnetic complexes are *shifted* more than on half of their linear size regarding each other. Besides, the FMC classification suggests two different cases to be distinguished by themselves: a new magnetic complex located about the leading *(p)* or following *(f)* part of another old one arises either on the part of a pole (*class IIa*) or on the part of solar equator (*class IIb*) |
| III | 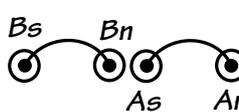 | A new magnetic complex arises *beside* another old one *without* any significant displacement *on latitude* i.e. a new magnetic flux rope is located either directly *behind* the following *(f)* part or (less often) *in front* of the leading *(p)* part of an old rope |
| IV | 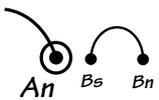 | This dynamic class is a variant of the FMC class *III* which has to do with two magnetic complexes of intrinsically *various sizes*. Besides, two different cases should be told apart as subclasses: either a new smaller complex arises in front of the leading *(p)* part of a *greater* old one (*class IVa*) or a return situation takes place: a new greater complex arises in front of a *smaller* old one (*class IVb*) |
| V | 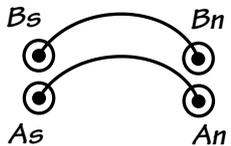 | A new magnetic complex is located *beside* another old one *without* any significant displacement *on longitude*. Besides, it is also essential to distinguish between two different cases: an emergence of a new complex takes place either on the part of a pole (*class Va*) or on the part of solar equator (*class Vb*) |
| B | 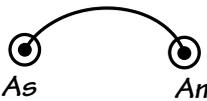 | An elementary magnetic configuration formed by *a unique* magnetic complex. In practice this situation mainly corresponds to a case of cleanly bipolar sunspot group. However, any trivial case of an isolated *unipolar* formation on the photosphere is also formally attributed by the FMC classification just to this class by definition |

Actually, if not to consider too difficult cases which were not confidently classified the analysis of a question on how observable flare importance depends on an FMC class has allowed to draw a number of the important statistical conclusions. In the present work these conclusions are summarized in Table 2 where the generalization of early found statistics (see Ikhsanov 1982, Table 1 and Table 2 therein) is represented. Looking at this data someone can find a number of obvious regularities.

(i) An overwhelming majority (almost *90 %*) of most powerful flares of an optical importance *3* with a reliably registered proton effect always occurs within those active regions which the topology closely corresponds to the class *I* and *II*, not to others ones. Besides, about *70 %* of them fit either the "clean" class *I* or a topological combination of the classes *I+II*, and just about *30 %* can be attributed to the "clean" class *II*.

(ii) Among all active regions where proton flares of an optical importance *2* have been detected, the overall contribution of magnetic configurations classified as *I* and *II* comes approximately to *80 %* but only a part of about *20 %* is provided by the classes *III–V*.

(iii) The distribution of non-proton flares through different classes maintains all general tendencies but the quantitative situation is a little bit other: here an overall part of most powerful flares of an optical importance *3* calculated through the classes *I-II* is already less than *60 %* (unlike *90 %* of a "proton" case); less powerful flares of an optical importance *2* also show significant decrease of the frequency of their registration within the classes *I* and *II*, namely about *45 %* (unlike *70 %* of a "proton" case), whereas their overall part calculated through the classes *III–V* accordingly grows.

(iv) The registration frequency maximum of more weak flares of an optical importance *1* and nanoflares (they are not submitted in Table 2, see instead Ikhsanov 1982, Table 1 and Table 2 therein) already shows very appreciable displacement to FMC classes *III–V*.

(v) In most cases, the active regions conforming with cleanly bipolar magnetic configurations of the class *B* seem to be able to produce weak flares only: here an overall part of flare events of an optical importance *2* is only about *11 %* and does not exceed *2–3 %* for more powerful flares which an importance is higher than *2*.

Therefore, the most flare effective are configurations of FMC classes *I*, *II* and *III* that is why sometimes we shall name them as "configurations of strong type". It is very indicative that while providing an independent consideration of those AR field structures which magnetic topologies should be numbered among "strong" FMC classes *I* and *II* according to Ikhsanov's early research (1982), Zirin and Liggett (1987) have later come just to a similar conclusion. Owing to this result we should emphasize the essentially important fact that even in most difficult *"combined"* cases the flare event of *maximum* importance seem to be always determined as a rule by the *general* FMC class (i.e. the *most expressed* among others, *not* the *sporadic* one) as it was also confidently shown (Ikhsanov 1982).

Physical reasons resulting in evident gravitation of the largest flares towards magnetic structures of FMC classes *I–III* appear to be indirectly connected to the fact that the configurations just of these three basic classes correspond much more than the other ones with the empirical criteria raising AR flare productivity as found in work (Ikhsanov 1982). Namely, in each of "strong" configurations each new magnetic bipole has two following features as distinct from "weak" ones: (i) as required, it rises inside (by "breaching" a field) or in immediate proximity from already existing complexes and (ii) by virtue of the "strong" FMC geometry it has an opportunity of the maximum effective (by "constricting" a field) interaction with another. At the same time neither of named conditions is coming out so far within more "weak" configurations, such as for those ones which have magnetic topology of classes *IV* and *V* (see Table 1).

**Table 2.** Observable frequency of powerful flare events (%) via registered optical importances in H$\alpha$ line and corresponding FMC classes (see details in work Ikhsanov 1982).

| Flare Importance | Proton flares (%) | | | | Nonproton flares (%) | | | |
|---|---|---|---|---|---|---|---|---|
| | I&II | II | III–V | B | I&II | II | III–V | B |
| 3F–3B | 62 | 26 | 12 | 0 | 24 | 34 | 40 | 02 |
| 2F–2B | 64 | 14 | 22 | 0 | 26 | 20 | 43 | 11 |

By no means, any complex system of interacting magnetic complexes initially forms in *subphotospheric* layers of the Sun. However, the magnetic energy saved up by such a system is next realized as flares and mass ejection within the *external* layers of the solar atmosphere where a range of most favorable physical conditions for this purpose is always presented. The closely established fact that the majority of powerful and, especially, proton flares always gravitates towards FMC classes *I* and *II* (see Table 2) is the most interesting point just in this context. For example, more later researches of so-called *white flares* have suddenly shown (Ikhsanov & Peregud 1988) that last ones tend to arise more often within magnetic configurations conforming with FMC classes such as *II* and *III*, not such as *I*. Just the simple generalization of these two results convincingly testifies to the assumption that physical conditions causing flare triggering within real magnetic configurations of various topology can be essentially different too.

### 2.5. Basic statements of IMC topological model

Taking as a basis the FMC dynamic classification, Ikhsanov (1985) has offered the rather simple topological model dealing with interacting magnetic complexes (hereafter the IMC model). At the same time an empirical technique allowing to analyse flaring configurations from the point of view of this model was also developed. Further, this technique has become the subject of testing and was repeatedly proved correct on extensive observational data in subsequent researches (Ikhsanov & Schegoleva 1980b, 1984; Ikhsanov & Peregud 1988; Ikhsanov & Marushin 1996a, 1996b, 1996c, 1998a, 1998b, 2000) where a lot of complex magnetic configurations was examined against the FMC model and the majority of registered phenomena was found to be naturally conforming with.

Actually the following basic statements seem to be observable results of repeatedly proven reliability and now form the basis of IMC model.

(i) The development of AR magnetic field occurs as a result of the consecutive or simultaneous rise of two or more bipolar magnetic ropes (so-called *magnetic complexes*) from under the photosphere which then prove themselves as new sunspot complexes inside or close by already existing sunspot group. The process of observable formation of several magnetic complexes is *a natural* development stage of any sunspot group, as well as the accompanying flare activity. The topology of interaction of two or more magnetic complexes closely determines the general magnetic field structure within the flare-productive active region by forming so-called *flaring magnetic configurations* (FMC).

(ii) In most cases each registered flare which an optical importance is not lower than 2 corresponds with the emergence of a new (i.e. earlier *not* marked) magnetic flow from under the photosphere. That is more, the most distinctive feature of any typical AR with powerful flares is an interaction of *at least two* such flows. Besides, an occurrence of a new magnetic complex on the photosphere is accompanied by a discrete series of flares from which an event of the *maximum* importance is determined by basic (i.e the *most* impressed) FMC topology. As a rule, the problem to reveal a new magnetic complex from within the existing group turns out to be carried out directly on the day or a day *prior* to flare.

(iii) The formalism of the FMC dynamic classification essentially allows to simplify a challenge of the topological description of magnetic configurations most frequently observable in real flaring ARs as well as a problem to forecast their flare productivity. The general regularity is that *most powerful* flares (in particular, *proton* ones) show an obvious tendency to group towards FMC classes *I–II*, whereas by reduction of flare importances the frequency maximum of their registration appears to be displaced to FMC classes *III–V*. Within the limits of any *fixed* FMC class the power of arising flares depends first of all on absolute and relative *capacities* of old and new magnetic complexes, as well as on geometrical *distances* between them.

(iv) The physical precedent ensuring flare situation development in the top layers of the solar atmosphere is created by gradual complication of AR magnetic field structure at the photospheric level. The last one occurs as a result of non-trivial interaction of *two or more* magnetic complexes which have emerged from under the photosphere. The distinctive feature of this process is that it takes place under the conditions of quite fixed magnetic field topologies which are essentially determined by the FMC class.

(v) An interaction of magnetic complexes is a necessary condition for the FMC formation, whereas the effects of a surplus tension of magnetic force lines influenced by topological structure of this or that configuration seem to be one of possible conditions ensuring the accumulation of surplus magnetic energy and its subsequent realization in the form of flare phenomenon.

(vi) The reduction or just as much the complete disappearance of mutual interaction between magnetic complexes (perhaps as a result of the rapid reconnection of certain force lines or by other known physical reasons causing magnetic field relaxation) leads to significant simplification of the general FMC structure, whereas *ingenious* topological features of that one seem to be able to determine observable specificity of a lot of the secondary effects accompanying flare phenomenon.

## 3. Applying the IMC model: general outlines

Generally, the minimum necessary data just allowing the FMC structure of a certain active region to be confidently reconstructed and investigated should include a lot of the important components. The first among them is a trustworthy information about the priority of magnetic flux emergence from under the photosphere along with the detailed data on polarity distribution at the photospheric level. Other things which should be also well understood are observably proved layouts determining an organization of magnetic flux ropes into interacting complexes. First of all this should mean an accurate knowledge of those field sources which footpoints of certain magnetic tubes are associated with.

In order to avoid any recurrences during the analysis of those specific flaring regions which will further appear in the second part of this work we shall only consider herein the general outlines of qualitative reasoning within the framework of the IMC model. We shall show that the topological formalism of a certain FMC class enables us to provide the reasonable interpretation for a lot of the observed phenomena, including the both obvious regularities and evolutionary effects carrying individual character. It frequently appears possible to ensure a consistent explanation not only to specific features of horizontal velocity field $v_\perp$ or to the facts of non-uniform sunspot development but also to certain phenomena influenced by irregular variations of the longitudinal $B_\parallel$ and transverse $B_\perp$ magnetic field. So, for example, the questions which can be clarified by such a way are just abnormal ratios between umbra and penumbra capacities and, specially, the reasons causing $\delta$-configuration formation. Here we assume those cases to be classified as "abnormal" where an umbra and penumbra surplus or just as well their deficiency unlike the "normal" values given by various models of sunspot magnetic field is observed (see, for example: Ikhsanov 1972, 1973).

Without any restriction of a generality the consideration of all typical outlines can be carried out on the most vivid example of the FMC class *I* which the formation usually accompanies the most powerful flares (Ikhsanov 1982). While analysing other FMC classes it is rather obvious that offered general outlines will stay fair to within the details connected to specific topology of appropriate configurations, mainly to the primary direction of mutual interaction of magnetic complexes. Therefore, as a typical illustration we shall consider the active region McMath 8362 (SD 093) of July 1966. There the specified topology closely corresponds just to a class *I* and hence all main effects at the photospheric level seem to be qualitatively explained by an interaction such as *"crosswise overlapping"* which two magnetic complexes $(A_N–A_S)$ and $(B_N–B_S)$ are involved in during their rise from under the photosphere (Fig. 2).

### 3.1. Magnetic hills: the specificity of mutual arrangement

Following the IMC model the top complex $(B_N - B_S)$ arisen earlier and the bottom one $(A_N - A_S)$ arisen later should form, as required, two pairs of magnetic hills close by photospheric sites where their footpoints are disposed. The corresponding nucleus of main sunspots at the points of their intersection with the photosphere are shown on Fig. 1. We shall note that the distinctive crosswise situation of transverse magnetic field $B_\perp$ is observed on magnetograms (Zvereva & Severny 1970) close by the central area where magnetic flux ropes of these two com-plexes cross each other, rather the picture of penumbra filaments (Steshenko 1969) is quite characteristic too. The model assumes that the observed bifurcation phenomenon of transverse field $B_\perp$ lines is the natural result of the spatial imposing of two different magnetic flux ropes one onto another.

First of all the FMC model makes it easy to understand (Fig. 2) the reason due to which the magnetographic data allows us to observe *four* magnetic hills *all together*. Really, we can find the *both* magnetic ropes *simultaneously* revealed, whereas they are disposed one *under* another and the magnetosensitive line Fe I $\lambda\, 6103$ Å in which the structure of the field $B_\parallel$ is studied forms itself within quite *thin* layer which the thickness is just about 100 km. Following the FMC model, magnetic force lines of the bottom complex $(A_N - A_S)$ are forced to exist in the situation of a limited freedom of their development. Therefore, while attempting to continue their rise towards the top layers of solar atmosphere they are necessarily *bent* on edges of the top complex $(B_N - B_S)$ where the emergence is still possible, and further form the topological "saddle" close by their symmetry centre. As a result four hills of a magnetic field are simultaneously presented at the level of Fe I $\lambda\, 6103$ Å line formation and this causes the characteristic situation of longitudinal field $B_\parallel$.

Obviously, the certain topological restrictions should be imposed on each magnetic configuration to maintain an effective compression of one magnetic field with another. The characteristic magnetic *"trap"* could play this role by allowing to compress a field up to the certain limits and, thus, to create the certain gradient $\Delta B_\parallel / \Delta r$. As appear, it is possible to assume in this connection that the elementary type of AR magnetic configuration conforming with the class *I* or its various modifications should be considered one of the most widespread kinds of a similar "trap" resulting in most powerful flares according to Ikhsanov's early statistics (1982).

Thus, we can find that the topology of a class *I* ("crosswise overlapping") obviously brings in its train a typical observed arrangement of magnetic hills and the well-known characteristic situation as two sunspot "chains" of opposite polarity resisting each other. Of course, many well known observational effects such as the shape of a "neutral" line $H_\parallel = 0$ as well as its significant inclination to the direction $N-S$, no less than its very typical shape seen at the edges of the active region which determines the characteristic *S*-shaped form of flare luminescence in H$\alpha$ line can also get by this way their most natural explanation.

### 3.2. Magnetographic variations of $B_\parallel$ and $B_\perp$ strengths

Magnetographic observations frequently show (see, for example: Gopasuk et al. 1963; Avignon et al. 1964, 1966; Godovnikov et al. 1964; Zvereva & Severny 1970) that the increase of the vertical gradient $\Delta B_\parallel / \Delta r$ along with its subsequent return back to the normal value is detected before flares. But just the topological membership of magnetic complexes of an active region to this or that FMC class should inevitably result in quite determined changes of magnetic field strength. So, for example, it is easy to show especially for the FMC class *I* that the phenomenon of vertical field $B_\parallel$ increase really *must* be observed close by sunspots of the complex $(B_N - B_S)$. Whereas at the same time its decrease above the sunspots of the complex $(A_N - A_S)$ with the suitable growth of transverse component $B_\perp$ strength *must* be also found.

Really, since local gradients of a vertical field $B_\parallel$ are proportional to $\Delta B_\parallel / \Delta r$ the IMC model predicts that their values should increase just for two magnetic hills of the top complex $(B_N - B_S)$ as they should be affected by a tension turned *upwards*, therefore their force lines should get more *vertical* direction (Fig. 3). On the contrary, the same increase measured for two magnetic hills of the bottom complex $(A_N - A_S)$, naturally, should be much less because of the opposite reason. The observation of longitudinal field gradients made for the active region McMath 8362 (SD 093) (Zvereva & Severny 1970) really shows the significant intensity growth of longitudinal field $B_\parallel$ in the site between two hills of the top complex $(B_N - B_S)$ and the much smaller one (or just its full absence) close by hills of the bottom one $(A_N - A_S)$. More accurate measurements made in the same work simultaneously at two different levels corresponding to the heights of $\lambda\, 5250$ Å and $\lambda\, 6103$ Å magnetosensitive line formation have also shown that (i) the vertical field $B_\parallel$ changes much more smoothly at the top level and (ii) it is twice less than on the bottom one by its absolute value. According to the IMC model this fact should mean that magnetic field structures are much *stronger* "squeezed" to the photospheric plane just in *deeper* layers as confirms again the hypothesis assuming the presence of some *additional* pressure being exerted upon the top complex from *below*.

Especially we have to underline the fact that with other things being equal the relative capacities of interacting complexes should essentially affect the ratio of $B_\parallel$ and $B_\perp$ observable intensities within the framework of the IMC model. Really, any situation when the capacity of the bottom complex $(A_N - A_S)$ essentially exceeds that one of the top complex $(B_N - B_S)$ should provide the more effective "superseding" of both magnetic flux ropes upwards and this will result, as required, in the greater observable intensity increase of the component $B_\parallel$ in comparison with the component $B_\perp$. On the contrary, when the top complex $(B_N - B_S)$ is more powerful than the bottom one, the specified "superseding" should be much less significant, and so magnetic force lines of the bottom complex $(A_N - A_S)$ appear more strongly "squeezed" to the photospheric plane. In its turn this will immediately create a return situation: an intensity increase of transverse magnetic field $B_\perp$ in comparison with the vertical field $B_\parallel$.

Following the IMC model, all similar distinctions in interaction intensity of magnetic complexes should just as well result in so various capacity of observed flares. It is useful to note in this sence that exactly this effect is *extremely* characteristic for *complex* active regions (Ikhsanov 1982) where much more vast horizons lie before magnetic field structures as regards the variety of their original topology and possible ways of their non-trivial preflare reorganization.

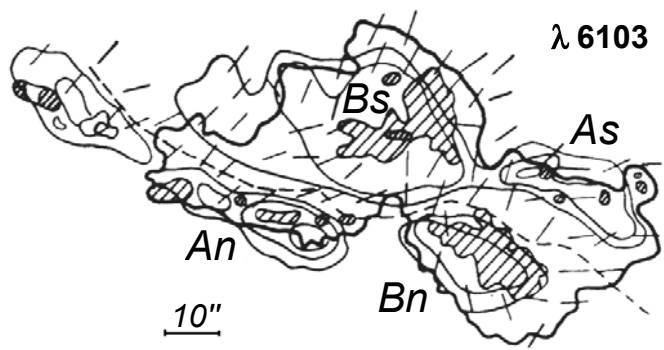 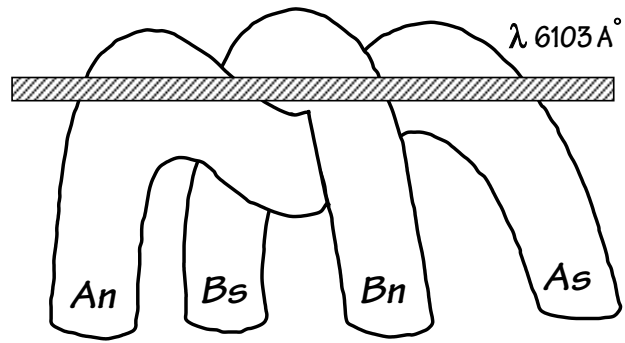

**Fig. 1.** The drawing shows the most typical photospheric arrangement of four "hills" of longitudinal magnetic field $B_\parallel$ on the image-plane magnetogram recorded in Fe I $\lambda$ 6103 Å line. The observable "quadripole" situation is demonstrated on the example of McMath 8362 (SD 093) flaring region. Thin closed curves mark the distribution of $B_\parallel$ contour levels, the directions of horizontal field $B_\perp$ are represented by long strokes. The solid shading and curves show sunspot penumbras and the common group penumbra respectively, dashed curve marks the shape and direction of the polarity inversion line $H_\parallel = 0$. The *bifurcation* phenomenon of transverse magnetic field $B_\perp$ directions should be also well observed.

**Fig. 2.** Schematic picture of *3D* interaction topology for two magnetic complexes $(A_N - A_S)$ and $(B_N - B_S)$ within the configuration such as *"3D crosswise overlapping"* (Ikhsanov 1974, 1982) is drawn. This simple model qualitatively explains not only *the bifurcation* phenomenon of transverse field $B_\perp$ directions but also the specific arrangement of four "hills" of longitudinal field $B_\parallel$. The spacial layer of Fe I $\lambda$ 6103 Å magnetosensitive line formation is shown by shading. Within the framework of such a topology the *"saddle"* symmetry centre approximately corresponds to the geometrical centre of an active region like shown on Fig. 1.

### 3.3. The deficiency, surplus and asymmetry of sunspot umbra and penumbra

Well-known that direct magnetographic data is not at all the only source from where irregular variations of magnetic field components $B_\parallel$ and $B_\perp$ can be successfully obtained. This phenomenon should also manifest itself in observable features of the umbra and penumbra of appropriate sunspots (i.e. the shape, position and capacity). But the deficiency of an internal (an external respectively) penumbra concerning main sunspots of the top complex $(B_N - B_S)$ (the bottom one $(A_N - A_S)$ respectively) is also well explicable by the IMC model (Fig. 3). Really, it is the most natural to connect it with the aspiration of the "feet" of magnetic arches to more (or less respectively) vertical direction. That seems to be an evident result of the decrease (or increase respectively) of an external "squeeze" from *above* i.e. on the part of old magnetic arches. At the same time the return of early affected penumbra picture to its normal state usually observable after the flare should be explained by the fact that the majority of parasitic deformations *must* significantly decrease after the contact area of two magnetic flux ropes is partially broken by flare and so the natural rise of the both field structures towards the upper atmospheric layers becomes available again.

So, for example, while analysing crimean images made by Steshenko (1969) we can conclude that the observable sunspot formation in the northwest part of the active region McMath 8362 (SD 093) went in the unusual way. Exactly, the penumbra has appeared first on 5–6 July, whereas the umbra has arisen later on 7 July, moreover the powerful $\delta$-configuration was simultaneously observed within the region. But according to the IMC model the considered site of the region concerns just to the bottom complex $(A_N - A_S)$ which magnetic flux ropes "squeeze out" the complex $(B_N - B_S)$ from *below* and so cannot freely develop themselves as a result of strong counteractions on the side of its magnetic force lines. Thus, there is nothing curious in the fact that the both sunspots of the bottom complex $(A_N - A_S)$ have just *the penumbra* at the initial stage of the active region evolution. Really, as its suitable force lines appear inclined ("squeezed") to the photospheric plane, their inclination angle is much great than it is necessary for a normal umbra formation. It is very typical in this connection that in further rise of the bottom complex $(A_N - A_S)$ its force lines became more vertical and compact on the edges, so that this has entailed on 7 July the high-grade umbra formation around main sunspots. As was to be expected, the final formation of sunspots as well as their detachment from the common $\delta$-configuration penumbra has taken place *only after* the magnetic field of the bottom complex $(A_N - A_S)$ has spatially stood apart as a result of topology simplification.

It is quite obvious that some reorganization of AR magnetic field is necessary under the conditions of a "trap" to realize such a spatial isolation. Really, early observations made in Crimea (see Steshenko 1969, Fig.4 herein) show that such a reorganization took place in the period 6–7 July, that is simultaneously with the registration of proton flare. What's more, the reorganization of a field is also proved correct by Crimean magnetograms on 5–7 July (see Zvereva & Severny 1970, Fig.4–6 herein). Actually, the maps of transverse field $B_\perp$ made on 5 and 6 July show the bifurcation of suitable force lines *just* within the area where an "overlapping" of magnetic flux ropes concerning the complexes $(A_N - A_S)$ and $(B_N - B_S)$ has to occur in accordance with the IMC model. But already on 7 July these lines are closely oriented in such a manner that is very distinctive for the *bottom* complex only, that is in the direction from the area $A_N$ towards the area $A_S$. Following the IMC model this situation should mean the fact that magnetic field structures of the top complex $(B_N - B_S)$ were "superseded" upwards and then replaced by the field of the bottom one $(A_N - A_S)$ approximately on the height where the line Fe I $\lambda$ 6103 Å is formed. It is very significant that the analysis of penumbra filament picture found on 5–7 July can immediately lead someone just to a similar conclusion.

Therefore it is possible to see that the interaction of magnetic complexes in the form of their mutual "squeeze" is *extremely* capable to result in the situations where two or more sunspot nucleuses of opposite polarity are located within the limits of an extensive common penumbra. In other words this mechanism allows $\delta$-configuration to be naturally formed and what's more its capacity seems to be depending on the intensity of such a "squeeze". According to the FMC model, the $\delta$-strusture origin is physically caused by a surplus inclination to a photospheric plane of magnetic force lines concerning that complex which is under the conditions of the limited development freedom. For this reason it is possible to approve that the observation of $\delta$-structure phenomenon within the complex active region (as a typical example of an abnormal penumbra picture around the sunspots of opposite polarities) is the direct evidence of the fact that magnetic fields of two or more complexes are in the state of their intensive preflare interaction, and so the surplus tension affecting magnetic force lines of flaring configuration remains significant.

For example, early Crimean observations (Steshenko 1969; Zvereva & Severny 1970) confidently show that the reorganization of magnetic force lines at the photospheric level along with the powerful flare was required before 7 July 1966 not only to initiate the high-grade sunspot formation of the bottom complex $(A_N - A_S)$ but also to provide the fragmentation of the common penumbra which has next resulted in $\delta$-structure simplification. It is quite obvious that such a reorganization would be possible just due to a partial disappearance of a "squeeze" on the part of AR magnetic field conforming with the top complex $(B_N - B_S)$. In its turn, such an opportunity would be created by this field as a result of its penetration into the upper layers of an atmosphere along with the reconnection of several force lines. Therefore, it is logical to consider the flare phenomenon of 7 July to be a consequence of magnetic flux emergence and its interaction in the form of two magnetic complexes. As we intend to stay within the IMC model frameworks, we can assume that further development of this process has resulted (probably owing to a reconnection) in the simplification of magnetic field topology due to a partial removal of those parasitic tensions and deformations which the configuration has accumulated before.

The similar phenomenon which is also well explained by the IMC model is widely known morphological effect of penumbra asymmetry that is the situation when several sunspot nucleuses have an abnormally small penumbra on the one side or just have not that one at all (Fig. 3). So, for example, Crimean data on 5 July shows that several nucleuses without any outer penumbra are observed on sunspot boundary (see Steshenko 1969, Fig.4 herein) in the southeast part of the active region McMath 8362 (SD 093). But according to the IMC model all these nucleuses are related to the area $A_N$ conforming with the bottom complex $(A_N - A_S)$ which magnetic flux ropes are "squeezed" from *above*, that is on the part of the top complex $(B_N - B_S)$. Therefore, those magnetic force lines which could form otherwise the high-grade *outer* penumbra of appropriate nucleuses are really "strapped" in practice towards the *inner* area of the active region where the "saddle" symmetry centre is disposed. As a result an inclination angle which they make with respect to a perpendicular to the photospheric plane is a bit less than it is necessary for penumbra formation.

It is rather typical in this connection that just after the powerful flare on 7 July when the magnetic configuration has become simpler and parasitic tension forces have decreased according the IMC model, the outer penumbra deficiency of sunspot nucleuses within the area $A_N$ has really ceased to be observed. The simple analysis of this fact in the aggregate with already considered effect of slowed down penumbra development in the northwest part of the bottom complex $(A_N - A_S)$ just can produce additional interesting results with reference to the last one. So, for example, this generalization allows us to draw more fine conclusion that the system of suitable magnetic flux ropes was not only "squeezed" but also a little bit *twisted* with an inclination to the west, so that the bottom complex $(A_N - A_S)$ was affected by the top one $(B_N - B_S)$ not strictly on the centre of symmetry.

### 3.4. Variation of magnetic fluxes $F_N$ and $F_S$

Crimean magnetographic observations (Zvereva & Severny 1970) show the perceptible growth of an overall magnetic flux $(F_N + F_S)$ before flares along with its subsequent falling down to former values. The well-known phenomenon of magnetic flux inequality that is the change of observable values $(F_S - F_N)$ in due course is also frequently marked. Following the IMC model the first effect should be related to the increase of an overall vertical field $B_\parallel$ of an active region McMath 8362 (SD 093) owing just to the same mechanism which provides an effective "superseding" of a field of the top complex $(B_N - B_S)$ by emerging magnetic structures of the bottom system $(A_N - A_S)$. In its turn, the second effect can be the natural result of the greater increase of a flux of one polarity $F_N$ in comparison with another $F_S$ or a return situation. It is easy to understand that such a phenomenon should inevitably occur during the rise of the bottom complex $(A_N - A_S)$ under the conditions of its asymmetric "squeeze" by magnetic structures of the top system $(B_N - B_S)$. We shall note that just this asymmetric interaction takes place in the majority of real cases.

### 3.5. Variation of magnetic energy $|W|$, $W_\parallel$ and $W_\perp$

On a level with the data concerning changes of magnetic field gradients, some important information on the interaction of magnetic complexes can be obtained from the separate estimation of magnetic energy contained in the longitudinal $W_\parallel$ and transverse $W_\perp$ magnetic field and, essentially, of their relationship with the overall magnetic energy $|B|^2 V/8\pi$ within the volume $V$ of an active region. So, for example, according to Crimean observations (Zvereva & Severny 1970), the energy of a longitudinal field $|B_\||^2 V/8\pi$ in the active region McMath 8362 (SD 093) has increased before the proton flare on 7 July from $(1 \div 2)\, 10^{32}$ erg up to $9\, 10^{32}$ erg since $4\overset{d}{.}3$ till $6\overset{d}{.}2$ July and then has fallen down to former value, besides the same situation was detected for changes of an overall energy $|W|$. Obviously, it is impossible to explain this phenomenon by the ordinary emergence of a new magnetic flux from under the photosphere as it was offered by Zvereva and Severny (1970) because exactly the *absolute* increase of an *overall* energy $|W|$ within the volume of the active region takes place, *not* just its simple *redistribution* in favour of longitudinal field energy $W_\|$ at the expense of that one of a transverse field $W_\perp$.

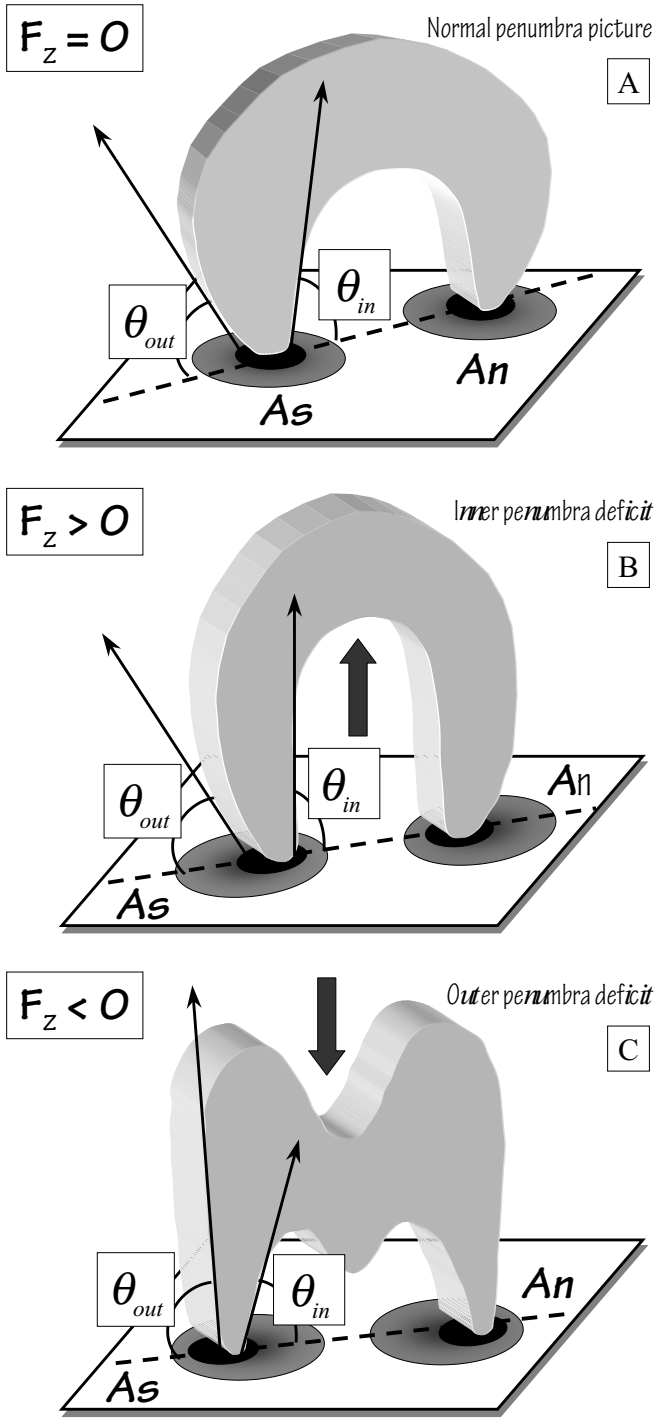

**Fig. 3.** Most typical alterations in observable sunspot morphology of a single magnetic complex $(A_\mathrm{N} - A_\mathrm{S})$ under the influence of an external force $\boldsymbol{F}$ are represented. As required, any foreing coercion deformes magnetic flux tube and has to change geometry of its force lines. Such a deformation immediately causes appearence of surplus angles $\theta_\mathrm{in}$ and $\theta_\mathrm{out}$ formed by the local direction of magnetic vector $\boldsymbol{B}$ with respect to the photospheric plane and so breaks their normal parity given by sunspot magnetic field models (see, for example: Ikhsanov 1972, 1973). As a result someone can detect either *(a)* a normal penumbra picture (there is no deforming force $\boldsymbol{F}$, so that $\theta_\mathrm{in} \simeq \theta_\mathrm{out} \ll \pi/2$ ) or *(b)* an inner penumbra deficit (deforming force $\boldsymbol{F}$ acts from below the magnetic arch, that is why $\theta_\mathrm{in} \to \pi/2$ ) or *(c)* an outer penumbra deficit (deforming force $\boldsymbol{F}$ acts from above the arch, and therefore $\theta_\mathrm{out} \to \pi/2$ )
.

However, the simple explanation of this phenomenon can be provided by the IMC model by assuming the fact that some parasitic "squeeze" takes place in addition on the part of the top complex $(B_\mathrm{N} - B_\mathrm{S})$ along with the emergence of the bottom one $(A_\mathrm{N} - A_\mathrm{S})$. Certainly, the specified energy decrease within the volume of the active region after the proton flare on 7 July can indicate the partial return of AR magnetic field *downwards* as it is supposed in work (Zvereva & Severny 1970). But the opposite explanation seems to be also possible and can be done with the IMC model. That one assumes the partial "breach" of the bottom complex $(A_\mathrm{N} - A_\mathrm{S})$ into the upper layers of an atmosphere of the Sun due to a reconnection of several force lines forming the common magnetic configuration $(A_\mathrm{N}-A_\mathrm{S}) + (B_\mathrm{N}-B_\mathrm{S})$. It is very interesting to note in this connection that at least two early established facts can be considered as good arguments in favour of the second assumption, namely (i) the vertical component of magnetic field $B_\|$ within the area where main sunspots of the bottom complex $(A_\mathrm{N}-A_\mathrm{S})$ are disposed is *increased* in due course, whereas those ones of the top complex $(B_\mathrm{N} - B_\mathrm{S})$ show signs of *fragmentation*; (ii) the distance between main sunspots of opposite polarity is also *increased* in due course for *the both* complexes, so that the further emergence of *the both* specified magnetic flux ropes gets its additional confirmation once more.

### 3.6. Morphology of chromospheric luminescence in H$\alpha$ line

In order to analyse the structure of AR magnetic field within the areas which are spatially brought nearer to the site of flare development it is necessary to consider a number of chromospheric phenomena observed in H$\alpha$ line. Really, it is shown by many authors (see, for example: 1964; Zirin & Lackner 1969; McIntosh 1972) that the structure of solar chromosphere observed in H$\alpha$ line can contains an important information on the structure of chromospheric magnetic field where direct observations are often too awkward. So, for example, bright sites seen through H$\alpha$ images spatially correspond to the areas with the prevalence of vertical field $B_\|$ whereas light and dark filaments indicate the sites where the domination of transverse field $B_\perp$ is usually found.

Unfortunately, there is no detailed H$\alpha$ data concerning the development of proton flare on 7 July in the active region McMath 8362 (SD 093) as the region was already near to the western edge of the limb. However, Crimean data on 6 July concerning the weaker flare of an optical importance *1* confidently shows (Zvereva & Severny 1970) that the flare brightening in H$\alpha$ line was observed along the *internal* side of magnetic hills concerning the top complex $(B_\mathrm{N}-B_\mathrm{S})$. Moreover, it was also found that H$\alpha$ emission of 7 July proton flare registered within the area of emphS-polarity was located from the *interior* of magnetic hill $B_S$. Summarizing the specified data it is possible to assume that H$\alpha$ emission has been detected close to those spatial areas where the top magnetic complex $(B_\mathrm{N}-B_\mathrm{S})$ adjoined the bottom one $(A_\mathrm{N}-A_\mathrm{S})$ according the IMC model.

It is well-known that flare brightening in H$\alpha$ line usually occurs in those areas of an active region where the sites of increased brightness were *already* marked before that is first of all in those areas where the intensity of a longitudinal field $B_\parallel$ is high enough. But under the conditions of the strong vertical field $B_\parallel$ and high conductivity of flare plasmas the reality of luminous substance movement *away* from a zero line $H_\parallel = 0$ directly along the *horizontal* plane at a level of the chromosphere seems to be doubtful as it would mean that the movement occurs *across* magnetic force lines of flaring configuration. On the contrary, the consideration of this fact along with the geometry of a typical magnetic arch should inevitably result in the conclusion that the phenomenon of H$\alpha$ luminescence reflects a situation when the exciting process gets its further spreading within the *vertical* plane. In other words, it appears to be going away from the site where the tops of interacting complexes $(A_\mathrm{N}-A_\mathrm{S})$ and $(B_\mathrm{N}-B_\mathrm{S})$ are contacting and next along the feet on either side of their magnetic arches downwards the areas disposed above the suitable footpoints in the chromosphere. Then the observable expansion of flare ribbons represents a usual orthogonal projection of a trajectory of exciting agent to the plane disposed in an atmosphere of the Sun at a height where H$\alpha$ line is formed.

Thus, it is logical to assume that observable expansion velocity of flare ribbons on either side of zero line $H_\parallel = 0$ is just the reflection of an actual velocity which is typical for the penetration of flare agent into the field structures of magnetic complexes $(A_\mathrm{N}-A_\mathrm{S})$ and $(B_\mathrm{N}-B_\mathrm{S})$. At last, we shall note that exactly such a way of interpretation of H$\alpha$ luminescence easily allows to explain the widely known situation where main sunspot nucleuses of an active region are surrounded by flare brightening as well as the distinctive emphS-like appearance of flare ribbons as that one is usually observed through H$\alpha$ images.

### 3.7. Sunspot fragmentation and special features of a horizontal velocity field $v_\perp$

Also well-known (see, for example: Gopasuk et al. 1963) that sunspot groups frequently show an effect of significant disintegration of large sunspot nucleuses by smaller ones that is the fragmentation and dissipation of an overall magnetic flux of an active region. This phenomenon usually finds out the close connection with flare activity, besides observable changes of sunspot planimetric features and overall sunspot number correlate with the moments of powerful flares. Within the framework of the IMC model the given correlation is natural for connecting with the acceleration of sunspot group disintegration as a result of partial removal of local parasitic deformations saved up within magnetic structures above sunspots. Really, it is quite obvious that such a removal should be one of the most natural consequences of powerful flares due to a fast reconnection of those force lines which conform with the suitable magnetic complexes. Moreover, well-known phenomena of the moderation and further acceleration of sunspot proper motions in preflare and postflare periods respectively can be well explained just by the same way. Actually, this is just due to a fast disappearance of some foreign influences like tension forces of magnetic lines forming flux ropes that those ones get some additional freedom and so can provide the observable velocity increase of those sunspots which are their footpoints. On the contrary, the specified preflare moderation of sunspot proper motions should correspond to the state when parasitic deformations are effectively accumulated as a result of a growth of foreign forces.

### 3.8. Certain individual regularities

At last it is necessary to note that certain individual regularities also get their qualitative explanation within the framework of the IMC model. So, for example, Palamarchuk (1973) showed that an interaction intensity of a close pair of sunspot groups appears to be depending on the *angle* which the primary direction of such an interaction (i.e. its projection on the photospheric plane) makes with bipole magnetic axes. With other things being equal, the regularity is that the angle is *smaller* the interaction is *stronger*. In work (Marushin 1998) we already paid attention to this result and found the latter to be quite conforming with the most common conclusions of work (Ikhsanov 1982) as applied to flares. Really, the FMC classification approves that just magnetic configurations of classes *Va* and *Vb* (where the specified angle is close to 90°) show on the average the *minimum* flare productivity among others. Meanwhile, the most *powerful* flares mainly gravitate either towards sunspot groups having FMC topology of classes *I* and *II* (where the maximum angle usually does not exceed 45° by virtue of geometrical reasons) or just as well to those ones of a class *III* (where the angle is always close to zero).

It is interesting to note that all topological structures of FMC classes *IVa* and *IVb* also have the specified angle extremely *small* just like structures of a class *III* but *never* produce powerful flares as a rule (see Table 2). Of course, this point can be easily considered as flagrant violation of early found reasonings. But we must not overlook the fact that *the capacity* of one magnetic complex regarding another is always extremely *small* for all members of this class (see Table 1). Being put into practice, this means that empirical flare productivity of such systems is always significantly *reduced* unlike the members of FMC class *III* where geometrical sizes are approximately identical. Thus, being summarized with the FMC general statements, this reasoning seems to be able to clarify a little bit the origin of Palamarchuk's empirical regularity by assuming that one to be partly influenced by special features of suitable magnetic topologies.

Therefore, despite the strong quatitativity of discussed IMC approach the range of its explanatory opportunities seems to be wide enough. Besides the basic advantage of reasoning within the framework of this model is that the latter logically allows us to draw a general conclusion from a lot of known observable phenomena as if they have the uniform physical reason. As shown above, to achieve this aim on a qualitative level it is often enough to examine an evolution of magnetic configurations within the active region of our interest as regards the formation of new magnetic complexes and their further interaction with those old magnetic field structures which have arisen earlier.

## 4. The IMC topological model: general discussion

Basic statements of the IMC model seem to be of utmost interest as regards two following tasks: (i) to find out the reasons and typical mechanisms of $\delta$-configuration formation within the complex active regions and (ii) to provide a consistent explanation for high flare productivity of such specific regions. The empirical relationship "$\delta$-configuration against powerful flares" marked by Künzel (1960) in the aggregate with the regularity "powerful flares against FMC classes *I–III*" established by Ikhsanov (1982) allows to presume the existence of an obvious genetic connection of $\delta$-structure phenomenon at a photospheric level on the one hand and three basic topologies of "strong" FMC classes *I–III* conforming with magnetic field structures above the photosphere on the other one.

### 4.1. Photospheric morphology and FMC topology

Adhering to the generally accepted electromagnetic notion of solar activity, it is most natural to assume that observable morphological features of any object registered in photospheric brightness field within the limits of an active region are mainly determined by the distribution of local magnetic fields. It is obvious in that case that any restrictions put onto this distribution by a certain class of magnetic configuration should be displayed themselves as a certain peculiarity of observable morphological picture. But the opposite statement seems also to be proved correct by many observations: any morphological phenomenon on the photosphere such as $\delta$-configuration can be able to maintain its long-living stage just under those conditions which allow restricted field structures to effectively preserve a set of vitally important parameters such as their general topology, their intensity, their local direction etc.

It is well demonstrated many times by photospheric observations that sunspots of opposite polarity *never* form any common penumbra in *clean* bipolar groups which are "ideal" cases conforming each one with a single pair of magnetic sources without additional complications. Really, the geometry of magnetic force lines of each single bipolar complex represents there just a simple *undeformed* arch system connecting magnetic areas of opposite polarity. Thus, there are no physical reasons to form the greater penumbra than it follows from a normal ratio of the umbra and penumbra capacities as provided by various models of sunspot magnetic field (see, for example:

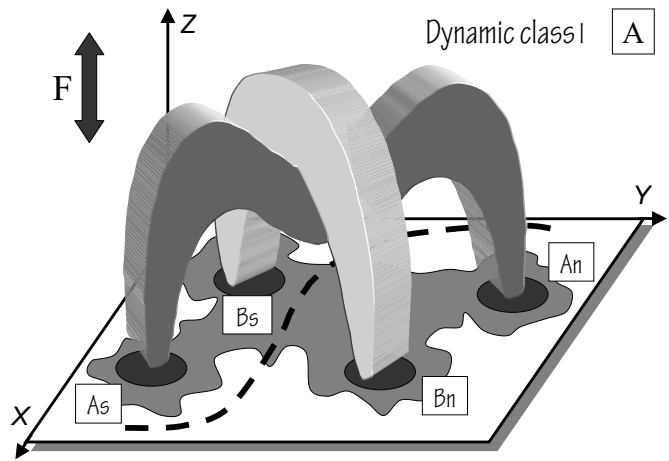

**Fig. 4.** Primary direction of deforming force $F$ in the presence of a pair of magnetic complexes $(A_N - A_S) + (B_N - B_S)$ within the framework of basic topology of dynamic class *I* ("vertical collision"). Each magnetic complex takes part in vertically oriented "top-by-top" interaction with another magnetic arch. This brings into action a number of evident topological constraints conditioned by vertical contracting deformations of both magnetic flux ropes approximately along the radial direction ($OZ$ axe). The explicite effect of *3D* crosswise overlapping of suitable magnetic force lines ("bifurcation") along with the closely distinctive penumbra peculiarity are the phenomena to be *always* created by such constraints in the first place among others (see also Fig. 5 and Fig. 6). Thus, such a morphology should be most naturally defined as $\delta$-structure of class *I*.

Ikhsanov 1972, 1973). Meanwhile it is necessary for the origin of high-grade sunspot nucleus (apart from other requirements such as intensity of a field about $1500$ G and a certain scale of magnetic flux ropes) that the inclination angle of magnetic vector $B$ to the photospheric normal does not exceed some appointed value (Ikhsanov 1972, 1973). Therefore, any *violations* of sunspot morphology should tend to occure just in those practical situations where the structure of magnetic field is fairly *complicated* and there are sufficient physical reasons for observable changes of a normal umbra / penumbra ratio. In other words such specific situations can be only conditioned by *foreign* forces and hence just *parasitic* tensions of magnetic flux ropes arisen as a result of mutual deformation of interacting complexes should be considered as suitable candidates to play this role as assumed by the IMC model, for example.

While finishing our consideration of the FMC classification it is necessary to pay attention to the following feature of three "basic" magnetic topologies *I–III* which seems to be conceptually important. It is easy to note at the close analysis of the whole system of formal classes offered by the FMC classification that *all three* spatial directions are represented there from the geometrical point of view without *any* exception, so that there is *exactly one* direction for *each* basic FMC class. Really, just three "basic" FMC classes are especially intended to formalize those practical situations where magnetic fluxes of existing field sources are drawn into topological interaction in conformity with the scheme *I* (vertical "squeeze" along the photospheric normal, Fig. 4), *II* (lateral "squeeze" along the direction $N-S$, Fig. 5) and *III* (lateral "squeeze" along the direction $E-W$, Fig. 6).

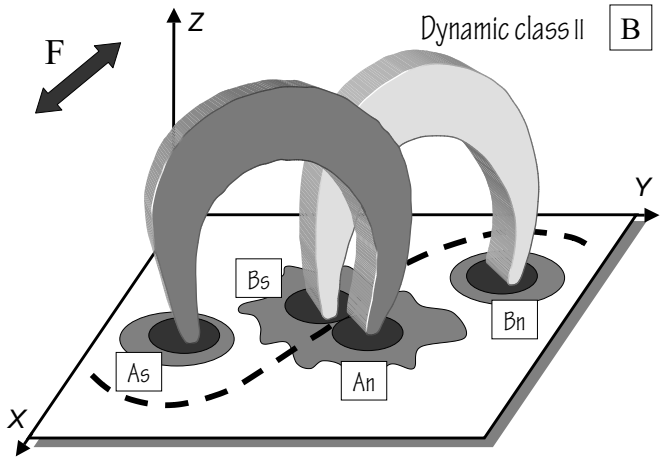

**Fig. 5.** Primary direction of deforming force $F$ in the presence of a pair of magnetic complexes $(A_N - A_S) + (B_N - B_S)$ when basic topology of dynamic class *II* ("*lateral collision*") is realized. Each magnetic complex is drawn into horizontally oriented "*side-by-side*" interaction with another magnetic arch which causes lateral contracting deformations of both magnetic flux tubes approximately along the direction $N-S$ ($OX$ axe). This type of interaction demonstrates just one more natural way (the *second* one, see also Fig. 4 and Fig. 6) by which strong topological constraints affecting further magnetic field evolution should be created. Just as shown on Fig. 4 and Fig. 6, by changing local geometry of magnetic force lines, such constraints have always to lead, as required, to the closely definite anomaly of penumbra morphology. Such a phenomenon should be defined as *δ-structure of class II*.

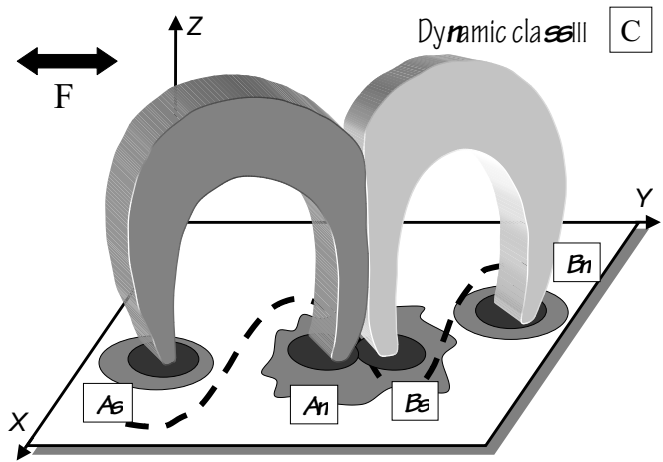

**Fig. 6.** Primary direction of deforming force $F$ in the presence of a pair of magnetic complexes $(A_N - A_S) + (B_N - B_S)$ under the conditions determined by basic topology of dynamic class *III* ("*frontal collision*"). Each magnetic complex is exposed in horizontally oriented "*foot-by-foot*" interaction with another magnetic arch which brings about lateral contracting deformations of their magnetic flux ropes approximately along the direction $E-W$ ($OY$ axe). Such a situation offers *a third* possibility by which another kind (see also Fig. 4 and Fig. 5) of topological constraints conditioned by this type of interaction should be initiated. Like Fig. 4 and Fig. 5, by affecting local magnetic field characteristics, such constraints have just as much to produce the closely definite irregularity of penumbra morphology. That one should be formally defined as *δ-structure of class III*.

In other words an attempt to examine any real *complex* configuration of AR magnetic field in order to find out the existence of topological structures of a class *I*, *II* or *III* and next to distinguish between them should be considered in a sense as a direct analogue of geometrical decomposition of a "perturbation" vector $F$ (i.e. resulting vector of tension forces of magnetic lines) into the basis formed by three spatial directions running along the axes $OX$, $OY$ and $OZ$.

By virtue of tridimentionality of AR physical space the existence of *three* basic FMC classes has basic value and seems to be a matter of principle. As appear, this state of things actually means that the FMC dynamic classification being considered as a consistent formal construction has in a sense the same geometrical property that is similar to an algebraic feature of the completeness of a basis constructed on three vectors in three-dimensional linear space. Being put into practice, this feature allows to expect that an attempt to formally construct any other topological classification of magnetic complexes based on *similar* principles or to expand the FMC classification by new *basic* classes will have at least one obvious consequence. Really, all such topologies should be derivatives from three basic FMC classes *I–III* and hence should be unequivocally described in early discussed terms of spatial interactions along three basic directions "bottom–top", $N$–$S$ and $E$–$W$ or some *combination* of those ones. This is just because no other linearly independent direction exists in three-dimensional space of an active region (Fig. 4, Fig. 5 and Fig. 6).

### 4.2. The δ-configuration phenomenon

While analyzing the FMC classification (Ikhsanov 1982) it is easy to understand that the certain topological restrictions should occur during the evolution exactly in those active regions where the configuration of a field meets requirements of discussed FMC classes *I–III*. Such restrictions should be realized by virtue of the configuration geometry and, in particular, they can affect the angle which force lines forming the "feet" of magnetic arches can make with an external normal $OZ$ to the photospheric plane. Thus, the local intensity of a vertical field $B_\parallel$ in the given area should essentially change itself and this should immediately affects observable sunspot parameters. So, for example, it was shown above that the observable surplus (or deficiency) of sunspot penumbra is capable to be effectively created owing to a "squeeze" (or a rise) of the suitable magnetic force lines with the respect to the photosphere.

It is easy to see that three different primary directions of an external deforming force $F$ can be distinguished between them in the presence of a pair of magnetic complexes $(A_N - A_S) + (B_N - B_S)$ depending on specific conditions of their interaction. These three directions are shown on Fig. 4, Fig. 5 and Fig. 6 along with three basic topologies *I–III* according to the FMC classification. It is also shown how the specificity of *3D* interaction of such two complexes can determine basic formative mechanisms for three different types of δ-structures in those real cases where FMC configurations of classes *I–III* are realized. On a level with the observable sunspot morphology, any strong variations of deforming force $F$ should considerably affect magnetic field characteristics and so have to render influence to observable flare features. By the same way this mechanism seems to be able to produce a number of specific *secondary* phenomena in the photosphere, the chromosphere and the corona.

On the one hand, it is possible to assume that three formal formative schemes answering for magnetic topologies of FMC classes *I–III* should *theoretically* conform with three *real* but *different* evolutionary ways which seem to be *equally* capable to result in the formation of $\delta$-configuration phenomenon. As shown above, the majority of observable effects accompanying the common penumbra formation is successfully interpreted by the IMC model in terms of magnetic fields affected by certain kinds of deformations during the interaction of magnetic complexes. In this sense it is *extremely* typical that exactly in those *real* active regions where the field topology precisely conforms with (or is highly close to) one of the "strong" formal FMC classes (*not* to *others* ones) the empirical precursor of most powerful flares ($\delta$-configuration ) is more often observed *in practice*.

On the other hand, numerous observations show that most *powerful* flares obviously gravitate just to those active regions which have their magnetic topology well conforming with FMC classes *I–III*. Besides, the IMC model qualitatively explains why it occurs. Really, in this work we consistently intend to adhere to working hypotheses (Ikhsanov 1982) that the physical reason (or at least *one* of many possible reasons) of solar flares is the interaction of magnetic complexes. While analysing a set of possible topologis of such complexes we have to recognize that *exactly* the configurations of FMC classes *I–III* ("crosswise overlapping", "lateral squeeze" etc) should most *effectively* create the situation where this interaction has *maximum* chances to appear the strongest.

The simple generalization of two last circumstances allows us to approve that (i) the formative mechanisms of $\delta$-configurations considered above are not only the most *natural* ones from the theoretical point of view but also are most likely *realized* in practice within the complex active regions with powerful flares and (ii) the known empirical feature of high flare productivity of active regions containing $\delta$-configurations (early pointed out by Künzel just as a simple statement) receives now the qualitative but quite strict (i.e. all discussed effects are due to the action of concrete forces) and consistent explanation (i.e. the majority of the reasonings based on the IMC model seems to be perfectly agreed with direct observations).

### 4.3. On the opportunity to classify $\delta$-configurations

In opinion of authors, the early specified genetic connection "$\delta$-configurations vs basic FMC classes" creates a logical precedent for the situation when the formal classification of $\delta$-configurations should be introduced into the use by analogy to the FMC dynamic classification. As required, this new classification should offer a similar system of formal classes with reference to various $\delta$-structures in order to provide an opportunity to differentiate those ones by their flare potential and to fall them into the structures of "strong" and "weak" type. In this case a new formal concept "$\delta$-*configuration of a class I, II, III*" should be understood by definition as "an observable morphological phenomenon on the photosphere where several sunspots of opposite polarity are spatially situated all together within the limits of a one common penumbra and magnetic flows of suitable field sources are topologically interacting under the circuit *I* (vertical "squeeze"), *II* (lateral "squeeze" along the direction $N-S$), *III* (lateral "squeeze" along the direction $E-W$)".

It is obvious that such a terminological approach should ensure the situation when the minimum information alredy allowing to explain many observable features of $\delta$-structure on a qualitative level (first of all, its morphological and flare characteristics) is *always* contained itself directly in the formal *name* of an object to be investigated. So, for example, by using the term such as "$\delta$-structure of a class *II*" someone would be allowed to form a primary notion of the way by which the corresponding AR is practically organized in outline and, what's more, would get a chance to do it just *without* any attempt *to look at* the object of his interest.

The authors assume this approach to be rather expedient and constructive just because it is capable to create a lot of obvious advantages for the further AR analysis *beyond* all dependences on its purpose and its specific character. Really, such a decision (i) allows to result in more *laconic* description of $\delta$-structure morphology *at* a photospheric level; (ii) unequivocally identifies *the topology* of mainly interacting magnetic flows *above* the photosphere and (iii) contains the minimum data need for the qualitative physical interpretation (and, what's more, for the forecasting) of a number of *secondary* phenomena registered (or still *expected* to be registered) within *the upper* layers of solar atmosphere such as the chromosphere and the corona. Namely, the sites and shapes of hydrogen emission in H$\alpha$ line, the probable localization of non-thermal $x$-ray sources and even the primary expected direction of the charged particle ejection could be numbered among them to begin with.

## 5. The IMC topological model and flare phenomenon

While considering the IMC model it is easy to notice that the majority of its general statements is constructed on the analysis of the whole complex of those phenomena which usually accompany the preparation and development of flare process from the existential point of view. The present work is the generalization of early made researches (Ikhsanov 1974, 1982, 1985; Ikhsanov & Schegoleva 1980a, 1980b, 1982, 1984; Ikhsanov & Peregud 1988; Ikhsanov & Marushin 1996a, 1996b, 1996c, 1998a, 1998b, 2000) and pursues the purpose to show that the application of the IMC model as means of empirical interpretation of observable phenomena in flaring active regions is rather fruitful.

It is *especially* important to note all along with the explanatory opportunities offered by the IMC model that the empirical reasoning within its framework does not mention in any way any subtleties of the concrete physical mechanisms initiating preflare reorganization of a field. Specifically, it *does not* enter into the conflict to the strict statements of modern theories offering various plasma instabilities which should develop in magnetized volume under certain conditions, next should create a physical precedent for the fast release of the "surplus" energy accumulated within the configuration and, hence, should become the physical reason of a flare. In particular, there are *no* contradictions with the well-known hypotheses approving that trigger mechanisms of a flare are probably determined by one of various MHD instabilities affecting the plasmas in a magnetic field or just as well by several thermal instabilities (such as the infringement of plasmas thermal balance because of the radiative energy loss or the dependence of its electric conductivity on a temperature), whereas several kinetic ("current") instabilities seem to be responsible for the rate of energy removal from the area of its initial release and, hence, for a number of "secondary" phenomena such as observable acceleration of the charged particles and the non-thermal radiation of flaring plasmas.

As a matter of fact, the flare phenomenon itself is actually considered within the framework of the IMC model just as a way of fast removal of local deformations (compression, twisting etc.) saved up by magnetic force lines within the topological peculiarities of a suitable FMC structure. Being discussed in highly exaggerated "mechanical" sense, the flare is represented here as a consequence of the most natural aspiration of a deformed physical system for the state with the smaller potential energy which is achievable by means of a non-elastic interaction. At the same time, the discussed effect of a fast FMC simplification corresponds within the frameworks of more strict "electromagnetic" interpretation to the fast reorganization of a magnetic field (probably owing to magnetic reconnection) and its further relaxation to the more potential configuration. The latter is as much as possible current-free one and, hence, it is more energetically natural: the energy of "surplus" currents should only increase the overall energy of any magnetic configuration. As a result of such a reorganization the complex magnetic configuration essentially becomes simpler, meanwhile the appropriate part of a "free" magnetic energy accumulated by suitable magnetic structures (i.e. its "surplus" considered with reference to the sheer potential field created by the completely identical distribution of photospheric sources) is released out and next observed as flare phenomenon.

## 6. Conclusion

By assuming as a basis the whole set of reliable empirical regularities of a kind "various magnetic topology vs various morphology, kinematics and flare productivity of an active region" the IMC model attempts to construct a logical generalization for all these regularities in their community and to consider those ones as having the uniform physical reason. For this purpose the IMC model tries to reformulate the named problem in terms of a kind "specific FMC class vs specific observable features" in order to demonstrate that physical conditions of flare situation development within magnetic structures of various FMC topologies are also essentially *various*. That is why, in opinion of the authors, the discussed model seems to be rather convenient tool in its *practical* relation. Really, in order to explain the specified regularities the IMC model provides the consistent qualitative theory. The latter considers several topological reasons which should initialize an accumulation of different energy reserves in different locations. For this purpose the model consecutively analyzes those different "parasitic" forces which the configuration is affected by and which seem to be able to produce different deformations of magnetic flux ropes, including "overcritical" ones.

Just with reference to this reasoning the very important question is whether the specified distinction of *conditions* can mean as well the distinction in *physics* of flare trigger mechanisms depending on a class of magnetic configuration or there is a certain *primary* mechanism in any case. In the last case the point of general importance is to understand to *what* extent this assumption can be correct if the answer is *yes*. In other words, the subject which is vitally important to be clarified is the question to what extent the concrete magnetic *topology* (i.e. the concrete FMC class) can determine real ways of practical realization of concrete theoretical *scenarios* offered by the modern theory of flares within the framework of standard models, including current-free ones.

Really, all *theoretical* scenarios offered by most "popular" models are quite equal in rights from the formal point of view. However, the question on their *practicability* under the conditions of a specific flaring region requires additional checking and still remains far from clear. It is obvious that the decision of this problem should include an accurate consideration of that specific role which magnetic configurations of various FMC classes can play in the preparation and development of flare situations according with suitable theoretical scenarios as well as the analysis of possible restrictions provided for the concrete FMC structure by concrete models of flares.

For example, the point of utmost interest is to revise new observational data against the question on how various power phenomena accompanying flares (such as their H$\alpha$ emission) spatially correspond with systems of vertical electric currents $J_z$ within the limits of an active region (see, for example: Canfield et al. 1991, 1993; Leka et al. 1993). There are just as much several specific subjects which could be also numbered among the most interesting questions, namely a role of zero points of AR magnetic field (see, for example: Demoulin et al. 1994) and the analysis of basic conclusions following from the topological model of vortex photospheric flows (Gorbachev & Somov 1988) just with reference to the IMC model.

It is completely obvious that an attempt to look for some evidences *pro* and *contra* several modern flare models could create a good opportunity to clear up the specified questions. All along with other observational tasks to be considered on this way, such an attempt should mean the phenomenological analysis of some specific *predictions* given by selected models as regards those morphological, kinematic and magnetic phenomena which get their *successful* explanation within the framework of the IMC model. The attempt of such an analysis could be based on all available data obtained from our consideration of a number of complex flaring regions and, hence, it should become the subject of our further researches.